# CL-MRI: Self-Supervised Contrastive Learning to Improve the Accuracy of Undersampled MRI Reconstruction

Mevan Ekanayake[a,b,*], Zhifeng Chen[a,c], Mehrtash Harandi[b], Gary Egan[a,d], and Zhaolin Chen[a,c]

[a]*Monash Biomedical Imaging, Monash University, Australia*
[b]*Department of Electrical and Computer Systems Engineering, Monash University, Australia*
[c]*Department of Data Science and AI, Monash University, Australia*
[d]*School of Psychological Sciences, Monash University, Australia*

## Abstract

In Magnetic Resonance Imaging (MRI), image acquisitions are often undersampled in the measurement domain to accelerate the scanning process, at the expense of image quality. However, image quality is a crucial factor that influences the accuracy of clinical diagnosis; hence, high-quality image reconstruction from undersampled measurements has been a key area of research. Recently, deep learning (DL) methods have emerged as the state-of-the-art for MRI reconstruction, typically involving deep neural networks to transform undersampled MRI images into high-quality MRI images through data-driven processes. Nevertheless, there is clear and significant room for improvement in undersampled DL MRI reconstruction to meet the high standards required for clinical diagnosis, in terms of eliminating aliasing artifacts and reducing image noise. In this paper, we introduce a self-supervised pretraining procedure using contrastive learning to improve the accuracy of undersampled DL MRI reconstruction. We use contrastive learning to transform the MRI image representations into a latent space that maximizes mutual information among different undersampled representations and optimizes the information content at the input of the downstream DL reconstruction models. Our experiments demonstrate improved reconstruction accuracy across a range of acceleration factors and datasets, both quantitatively and qualitatively. Furthermore, our extended experiments validate the proposed framework's robustness under adversarial conditions, such as measurement noise, different k-space sampling patterns, and pathological abnormalities, and also prove the transfer learning capabilities on MRI datasets with completely different anatomy. Additionally, we conducted experiments to visualize and analyze the properties of the proposed MRI contrastive learning latent space. Code available [here](#).

[*]Corresponding author.
E-mail address: [mevan.ekanayake@monash.edu](mailto:mevan.ekanayake@monash.edu)

# 1 Introduction

Magnetic Resonance Imaging (MRI) distinguishes itself among imaging technologies for its ability to provide superior contrast of soft tissues without ionizing radiation, offering safe and detailed insights into the body's internal structures (Nishimura, 2010). This makes it a preferred choice for diagnosing a broad spectrum of medical conditions related to human functions and anatomy, e.g., the musculoskeletal system, the brain and nervous system, and cancer-related illnesses (Grover et al., 2015). However, the extended scan times of MRI constitute a significant drawback, leading to discomfort for patients, motion artifacts in the images, and increased costs for resource allocation (Zbontar et al., 2019). These drawbacks are often mitigated by undersampling the MRI data in k-space. However, the undersampling inevitably leads to a deterioration in image quality, including aliasing artifacts, blurring, and loss of resolution, which can hinder clinical diagnoses using the images.

Over the past few decades, numerous techniques have been explored to reconstruct high-quality MRI images from those undersampled data and to speed up MRI (Wang et al., 2021). Initial advancements were achieved through parallel imaging methods, which utilize multiple receiver coils to capture spatial information simultaneously (Griswold et al., 2002; Pruessmann et al., 1999). These methods achieved reduced scan times owing to the unique spatial sensitivities of each coil. Subsequently, compressed sensing methods emerged for MRI which exploited the inherent sparsity of MRI images in a transform domain (Lustig et al., 2008). Together, these methods represent pivotal steps forward in MRI technology, offering faster, more patient-friendly scanning processes without compromising the detail and clarity essential for diagnostic imaging purposes (Zbontar et al., 2019).

Inspired by the surge in artificial intelligence, deep learning (DL) techniques have shown promising results for undersampled MRI reconstruction (Liang et al., 2020). These methods involve training a multilayer deep neural network to transform undersampled MRI data into high-quality MRI images, guided by a large collection of MRI datasets (Zbontar et al., 2019). The neural architectures of these models often include fully connected networks (FCNs), convolutional neural networks (CNNs), or, more recently, vision transformer networks (ViTs). Regardless of the architectural setup, DL methods for MRI reconstruction fall into two main categories (Wang et al., 2021): data-driven DL methods that convert low-quality, undersampled images into high-quality references (Ekanayake et al., 2024; Huang et al., 2019; Hyun et al., 2018; Ran et al., 2021; Sriram et al., 2020), and physics-constrained DL methods that iteratively solve the inverse problem of reconstruction (Aggarwal et al., 2019; Hammernik et al., 2018; Schlemper et al., 2017; yang et al., 2016; Zhang and Ghanem, 2018).

Despite the notable success of deep learning models in surpassing traditional reconstruction algorithms for MRI, their ability to generalize effectively across various conditions presents a notable challenge. This issue was highlighted during the second FastMRI challenge in 2021 (Muckley et al., 2021), with radiologists deeming all submitted 8X acceleration reconstructions as clinically unacceptable. Furthermore, a detailed review of deep learning applications in MRI reconstruction has indicated that current methods yield promising outcomes only with acceleration factors of up to 6X (Wang et al., 2021). Consequently, these observations emphasize the critical need for advancements in deep learning techniques for reconstructing undersampled MRI data, especially for improving the accuracy of the reconstructions. Given the rapid growth in deep learning research, there is clear and significant room for improvement in MRI reconstruction to meet the high standards required for clinical diagnosis (Chen et al., 2022).

One of the primary challenges faced by DL models is the insufficient presence of information at the input to obtain a desired output. Regardless of the neural network architecture, DL models fundamentally operate by identifying patterns from the input data, extracting the most relevant information in latent space, and using that latent space to obtain the desired output. The study by Tian et al., (2020) suggested that maximizing the task-relevant information at the input would yield high downstream task performance. In the context of undersampled MRI reconstruction, as the input data is undersampled to achieve higher accelerations, essential information crucial for the DL model to accurately reconstruct the output is significantly diminished, which leads to a decline in the reconstruction accuracy. This rationale is observed in almost all previous DL reconstruction papers (Ekanayake et al., 2024; Schlemper et al., 2017; Sriram et al., 2020), where the reconstruction accuracy has been reported to decrease with increasing acceleration factor.

To this end, we propose a novel self-supervised contrastive learning framework designed to enhance MRI reconstruction accuracy. Our framework, named Contrastive Learning MRI (CL-MRI), adapts the latent space representations of MRI images by maximizing mutual information among different undersampled representations and optimizing the information content at the input of the downstream DL reconstruction model. These enhanced latent representations serve as the base for image reconstruction, and our experiments show improved reconstruction accuracy, both quantitatively and qualitatively. Further investigations affirm our framework's robustness under adversarial scenarios, including measurement noise, different k-space sampling patterns, and pathological abnormalities. Moreover, CL-MRI proves to be compatible with transfer learning among MRI datasets with different anatomical regions, offering a critical solution to the prevalent issue of data scarcity in medical imaging. Additionally, we conduct detailed analyses of the contrastive learning MRI latent space, exploring the impact of contrastive learning for the downstream undersampled MRI reconstruction task.

Our paper is organized as follows: In Section 2, we discuss some fundamental background concepts on which our CL-MRI framework is built. In Section 3, we comprehensively describe our proposed framework, along with steps for flawless implementation. In Section 4, we present the list of our experiments followed by the results in Section 5. Sections 6 and 7 are dedicated to a comprehensive discussion and concluding remarks, respectively.

## 2 Background

This Section presents some background concepts that relate to the work covered in this paper.

### 2.1 Self-supervised Learning for Data Representation

Self-supervised learning, a subset of machine learning, has emerged as a powerful strategy for pretraining deep learning models, especially in scenarios where labels or ground truths are scarce or expensive to obtain (Liu et al., 2023). These approaches effectively utilize the input data itself to learn rich representations of data without the need for explicit external labels or ground truths. This is often accomplished by designing pretext tasks to understand its underlying patterns and features (Kolesnikov et al., 2019). Pretraining models using self-supervised learning have shown remarkable success across various domains, including natural language processing (NLP), computer vision, and audio processing (Newell and Deng, 2020).

In NLP, models like BERT and GPT have been pre-trained on vast text corpora using self-supervised tasks, such as predicting the next word in a sentence or filling in missing words, before being fine-tuned on smaller labeled datasets for specific tasks (Lan et al., 2020). Similarly, in computer vision, self-supervised pretraining helps models understand visual concepts and relationships within images (Newell and Deng, 2020) through tasks such as predicting one part of the data from another, reconstructing inputs, solving jigsaw puzzles of images, or employing contrastive learning techniques that aim to bring representations of similar data points closer together in the feature space while pushing representations of dissimilar data points further apart (Chen et al., 2020). These self-supervised pretraining techniques not only enhance model performance on downstream tasks but also improve generalization, making models more robust to variations in input data.

## 2.2 Contrastive Learning for Information Maximization

Contrastive Learning is a self-supervised learning technique, particularly within the field of computer vision, that aims to learn useful representations by contrasting positive pairs against negative pairs (Oord et al., 2019). This approach has gained significant attention for its effectiveness in learning features that are invariant to transformations.

In computer vision, Contrastive Learning revolves around the idea that similar, or "positive," pairs $(\mathbf{z}, \mathbf{z}^+)$ of data points (e.g., two different augmentations of the same image) should have closer representations in the latent space than dissimilar, or "negative," pairs $(\mathbf{z}, \mathbf{z}^-)$ of data points (e.g., augmentations of different images). This is achieved by defining a contrastive loss function, which is used to train a deep neural network. One of the most popular contrastive loss functions is the InfoNCE loss, defined as:

$$\ell_{\mathbf{z}, \mathbf{z}^+}^{\text{CL}} = -\log \frac{\exp(\text{sim}(\mathbf{z}, \mathbf{z}^+)/\tau)}{\exp(\text{sim}(\mathbf{z}, \mathbf{z}^+)/\tau) + \sum_{\forall(\mathbf{z}, \mathbf{z}^-)} \exp(\text{sim}(\mathbf{z}, \mathbf{z}^-)/\tau)} \tag{1}$$

for a positive pair of image representations. Here, $\tau$ is a scaling factor and $\text{sim}(\cdot,\cdot)$ is an arbitrary similarity metric. We refer readers to Chen et al., (2020) for more information on the InfoNCE loss function. This framework encourages the model to learn discriminative features by maximizing the similarity of positive pairs relative to negative pairs. Over time, this process results in a feature space where semantically similar examples are closer together, facilitating downstream tasks to become more generalizable to the variability in input data. For example, in a downstream classification task, contrastive learning ensures that the classification model is invariant to the augmentation in the input images.

## 2.3 DL MRI Reconstruction

In MRI, images inside a subject's body structure are captured using RF receiver coils in the frequency domain and are referred to as the k-space data. Let $\mathbf{x} \in \mathbb{C}^N$ represent a complex-valued MRI image and $\mathbf{y} \in \mathbb{C}^M$ represent the undersampled k-space. The goal of undersampled MRI reconstruction is to reconstruct $\mathbf{x}$ from $\mathbf{y}$. A direct reconstruction can be obtained by $\mathbf{x}_u = \mathbf{F}_u^H \mathbf{y}$, where $\mathbf{F}_u^H \in \mathbb{C}^{N \times M}$ represents the undersampled inverse Fourier encoding matrix, but $\mathbf{x}_u$ will contain aliasing artifacts, noise, blurring, and other image quality degradations.

The DL-based reconstruction problem is often formulated as the minimization of an objective function as below (Schlemper et al., 2018):

$$\hat{\mathbf{x}} = \underset{\mathbf{x},\theta}{\mathrm{argmin}}\{\|\mathbf{x} - f_{DL}(\mathbf{x}_u|\theta)\|_2^2 + \lambda\|\mathbf{F}_u\mathbf{x} - \mathbf{y}\|_2^2\} \qquad (2)$$

where the first term represents the DL regularization and the second term enforces data consistency (DC). Here, $f_{DL}$ represents the DL reconstruction model parametrized by $\theta$ and $\lambda$ is a tunable parameter. Wang et al. (2016) proposed the closed-form solution to the above optimization problem as $\hat{\mathbf{x}} = \mathbf{F}^H\hat{\mathbf{y}}$, where $\hat{\mathbf{y}}(k) = \begin{cases} \mathbf{y}_{DL}(k) & \text{if } k \notin \Omega \\ \frac{\mathbf{y}_{DL}(k)+\lambda\mathbf{y}(k)}{1+\lambda} & \text{if } k \in \Omega \end{cases}$. Here, $\mathbf{y}_{DL} = \mathbf{F}\mathbf{x}_{DL} = \mathbf{F}f_{DL}(\mathbf{x}_u|\theta)$ is the output k-space from the DL model. Thus, a cascade of DL models with interleaved DC blocks can be designed and the DL model parameters can be trained on a training dataset consisting of pairs of undersampled MRI images and the corresponding fully-sampled ground truth MRI images using a predefined loss function (Zbontar et al., 2019). Once the parameters are optimized, the final reconstruction can be obtained at the end of the final DC block of the cascade (Ekanayake et al., 2024).

In the above formulation of the MRI reconstruction task, the transformation from an undersampled image $\mathbf{x}_u$ to the desired ground truth image, $\mathbf{x}$ is a difficult task due to the large signal difference between the undersampled input and ground truth. In this work, we mitigate such difficulty by introducing a self-supervised pretraining step to map $\mathbf{x}_u$ to a contrastive learning latent space.

## 3 Methods

This Section presents our proposed framework for MRI Reconstruction and an overall schematic is shown in Fig. 1.

### 3.1 Contrastive Learning Pretraining

Motivated by the success of self-supervised contrastive learning, we propose to maximize the information content at the input of the downstream DL reconstruction model by mutually enhancing information contained in different acceleration factors. To achieve this, we undersample an MRI image in k-space by varying the acceleration factor, resulting in a collection of undersampled images derived from the same MRI scan, i.e., $\mathbf{x}_{u_1}, \mathbf{x}_{u_2}, \ldots, \mathbf{x}_{u_D} = \mathbf{F}_{u_1}^H\mathbf{y}, \mathbf{F}_{u_2}^H\mathbf{y}, \ldots, \mathbf{F}_{u_D}^H\mathbf{y}$ where $D$ is the total number of images generated from a single scan. Next, we extract latent space representations, $\mathbf{z}_1, \mathbf{z}_2, \ldots, \mathbf{z}_D$ from these undersampled images, i.e., $\mathbf{z} = T(\mathbf{x}_u|\phi)$, using a feature extraction model, $T: \mathbb{R}^{2N} \to \mathbb{R}^L$. Note that we represent complex-valued input images as two-channel real-valued input images, following common practice (Zbontar et al., 2019).

With a cohort of MRI scans, we apply contrastive learning by considering a pair of images generated from the same scan as a positive pair, and a pair of images generated from two different scans as a negative pair. We use the notation $(\mathbf{z}, \mathbf{z}^+)$ to denote a positive pair of representations and $(\mathbf{z}, \mathbf{z}^-)$ to denote a negative pair of representations. The rationale behind this design choice is to maximize mutual information among the images generated from a particular scan, thereby optimizing the information content at the input of the downstream reconstruction model. Our contrastive loss implementation for a positive pair of representations can be expressed as follows (Chen et al., 2020):

$$\ell_{\mathbf{z},\mathbf{z}^+}^{\text{CL-MRI}}(\phi) = -\log \frac{\exp(\text{sim}(\mathbf{z},\mathbf{z}^+)/\tau)}{\sum_{\forall(\mathbf{z},\mathbf{z}^+)} \exp(\text{sim}(\mathbf{z},\mathbf{z}^+)/\tau) + \sum_{\forall(\mathbf{z},\mathbf{z}^-)} \exp(\text{sim}(\mathbf{z},\mathbf{z}^-)/\tau)} \quad (3)$$

Here, $\text{sim}(\mathbf{u},\mathbf{v}) = \mathbf{u}^T\mathbf{v}/\|\mathbf{u}\|\|\mathbf{v}\|$ is the cosine similarity between two vectors and $\tau$ is a scaling factor that we set empirically. Note our implementation in Eq. (3) slightly differs from the conventional formulation in Eq. (1). In our implementation, the denominator contains the summation over the negative pairs as well as positive pairs, since we generate multiple accelerated images from a single scan, as opposed to just generating just two augmentations in conventional setting. The above loss is accumulated over the positive pairs of the entire dataset of generated images and used to train the parameters of the feature extraction network, $\phi$.

Closely observing the above loss function, the numerator incorporates the positive pair, whereas the denominator mainly includes negative pairs. This formulation enforces the 'pulling' of representations from the same scan towards each other (thereby maximizing mutual information) and the 'pushing' of representations from different scans (thus minimizing irrelevant information) away within the training batch. Furthermore, it has been demonstrated the above loss function, effectively maximizes a lower bound on mutual information (Oord et al., 2019). It is also important to note that this initial training phase is self-supervised, meaning it does not rely on fully sampled images. The original MRI measurements, **y** can be already undersampled.

### 3.2 Reconstruction using Contrastive Latent Features

Following the training of the feature extraction model, we can derive a contrastive latent representation for an undersampled image using $\hat{\mathbf{z}} = T(\mathbf{x}_u|\phi^*)$. These latent representations are then utilized to train a reconstruction model, employing the following loss function:

$$\mathcal{L}(\Theta) = \sum_{\forall(\hat{\mathbf{x}},\mathbf{x}_{gt})} \ell(\mathbf{x}_{gt}, \hat{\mathbf{x}}) = \sum_{\forall(\hat{\mathbf{z}},\mathbf{x}_{gt})} \ell(\mathbf{x}_{gt}, G(\hat{\mathbf{z}}|\Theta)) \quad (4)$$

Here $G: \mathbb{R}^{2N} \to \mathbb{R}^{2N}$ is the DL reconstruction model parametrized by $\Theta$. It should be noted that, contrary to traditional deep learning methods that use undersampled images, $\mathbf{x}_u$ as input for reconstruction models, our framework employs contrastive latent representations, $\hat{\mathbf{z}}$. After training, the trained reconstruction network can be applied to generate the final reconstruction of an undersampled MRI image, $\mathbf{x}^* = G(\hat{\mathbf{z}}|\Theta^*) = G(T(\mathbf{x}_u|\phi^*)|\Theta^*)$, where $\mathbf{x}^*$ is the final image.

### 3.3 Alignment and Uniformity of Contrastive Latent Features

Alignment and Uniformity are two key properties related to the contrastive loss which signifies the closeness of features from positive pairs, and the distribution of the normalized features on the latent space, respectively (Wang and Isola, 2020). Previous research has indicated that enhanced alignment and uniformity lead to improved efficiency in representing signals, consequently benefiting subsequent downstream tasks.

In the context of MRI, latent representations capture the most prevalent information among positive pairs while remaining unaffected by other noise factors such as aliasing artifacts, noise, and blurring. Within the proposed framework, alignment pertains to the similarity or closeness of contrastive learning features in the latent space derived from differently accelerated images of the same scan. It is defined on positive feature pairs as: $C_A \triangleq -\mathbb{E}_{\forall(z,z^+)}[\|z - z^+\|_2^\alpha]$, $\alpha > 0$, where $\alpha$ is a controllable parameter.

Uniformity measures how well the contrastive features are distributed in the latent space which is an indication of better generalization downstream tasks. It is often defined as the Gaussian potential on pairs of representations as: $C_U \triangleq \log \mathbb{E}_{\forall(z_i,z_j)}\left[\exp\left(-\beta \|z_i - z_j\|_2^2\right)\right]$, $\beta > 0$, where $\beta$ is a tunable parameter. In our experiment, we demonstrate that our proposed framework for MRI promotes Uniformity and Alignment properties which reinforces the gains in reconstruction performances further.

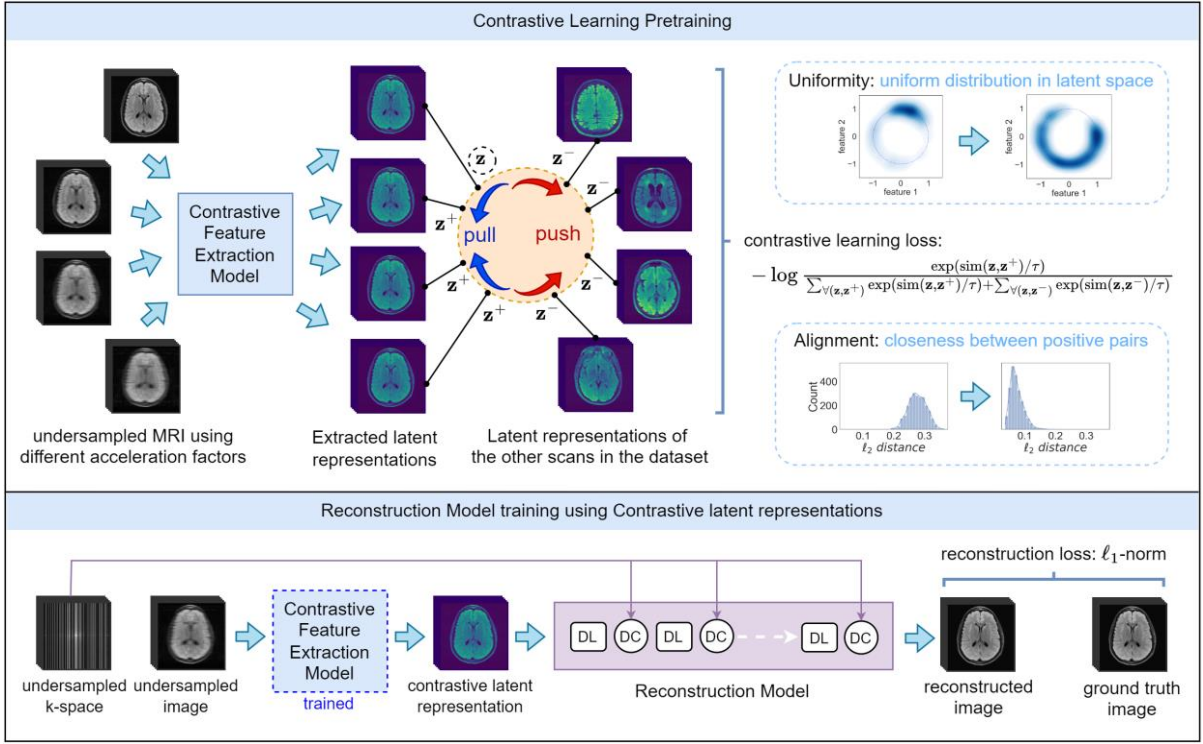

Fig. 1 Our proposed framework for self-supervised contrastive learning for undersampled MRI. Top panel: the contrastive learning pretraining phase. Bottom panel: the proposed DL reconstruction pipeline from the contrastive latent representations.

### 3.4 Implementation

Our reconstructions were conducted on a per-coil basis for multi-coil datasets and coil combinations were performed using the root-sum-of-squares method. To generate positive pairs, we applied undersampling to the k-space data using four distinct acceleration factors: 2X, 4X, 6X, and 8X. For the contrastive feature extraction network, we utilized the End-to-End Variational Network (E2E-VarNet) (Sriram et al., 2020). All experiments were conducted using an NVIDIA A100 GPU. Since processing the entire dataset at once is challenging due to the significant memory requirements, contrastive learning was implemented in mini-batches as shown in Algorithm 1. In our implementation, The E2E-VarNet was trained over 100 epochs with a minibatch size of 4, using an RMSProp optimizer with a learning rate set at 0.001. The pseudocode for the contrastive learning process of a sampled batch is shown in Algorithm 1.

---

**Algorithm 1: CL-MRI**

**Input:** batch size $B$, number of images generated from a single scan $D$, temperature scaling factor $\tau$, feature extraction model $T(\cdot \,|\phi)$

**for** sampled minibatch of MRI scans $\{\mathbf{y}_p\}_{p=1}^B$ **do**

  **for all** $p \in \{1, 2, \ldots, B\}$ **do**

    **for all** $q \in \{1, 2, \ldots, D\}$ **do**

      $\mathbf{x}_{u_{p,q}} = \mathbf{F}_{u_{p,q}}^H \mathbf{y}_p$

      $\mathbf{z}_{p,q} = T(\mathbf{x}_{u_{p,q}} | \phi)$

    **end for**

  **end for**

  We have a set of $DB$ number of latent representations extracted.

  Let $\mathbf{z}_i$ and $\mathbf{z}_j$ be an arbitrary pair of representations from this set.

  **for all** $i \in \{1, \ldots, DB\}$ and $j \in \{1, \ldots, DB\}$ **do**

    $s_{i,j} = \mathbf{z}_i^T \mathbf{z}_j / \|\mathbf{z}_i\| \|\mathbf{z}_j\|$

  **end for**

  **define** $\ell_{i,j}^{\text{CL-MRI}} = -\log \frac{\exp(s_{i,j}/\tau)}{\sum_{k=1}^{DB} \mathbb{1}_{[k \neq i]} \exp(\text{sim}(s_{i,k})/\tau)}$

  Here, $\mathbb{1}_{[k \neq i]} \in \{0,1\}$ is an indicator function evaluating to 1 iff $k \neq i$

  $\mathcal{L}(\phi) = \frac{1}{DB} \sum_{p=1}^B \sum_{q=1}^D [\ell_{p,q}^{\text{CLMRI}}]$

  update the parameters of $T(\cdot\,|\phi)$ to minimize $\mathcal{L}$

**end for**

**return** optimized feature extraction model $T(\cdot\,|\phi^*)$

---

## 3.5 Datasets

Our proposed framework was trained and validated using multi-coil FLAIR brain MRI data from the fastMRI brain dataset (Zbontar et al., 2019). This dataset comprises complex-valued k-space data captured using both 3T and 1.5T scanners using 16 receiver coils and features axial slices. For the out-of-distribution assessments in Experiment 2, only the reconstruction models were trained and validated using multi-coil proton density-weighted (PD) knee MRI data from the fastMRI knee dataset (Zbontar et al., 2019). This dataset comprises complex-valued k-space data captured using both 3T and 1.5T scanners using 15 receiver coils and features axial slices.

## 3.6 Undersampling

The undersampling was performed by retrospectively applying a binary mask to omit lines in the phase encoding direction of the fully sampled k-space. We adhere to the random undersampling protocol described by Zbontar et al., (2019), where each slice within a specific volume is subjected to the same undersampling mask while preserving a portion of the central frequencies.

## 3.7 Evaluation Metrics

To assess the reconstruction performance quantitatively, we employ two three widely recognized metrics: normalized mean square error (NMSE), peak signal-to-noise ratio (PSNR), and structural similarity index (SSIM) following the standard definitions in Zbontar et al., (2019). NMSE is an estimator of the overall deviations between reconstruction and fully-sampled ground truth. Lower values of NMSE indicate a better reconstruction. PSNR represents the ratio between the power of the maximum image intensity and the power of noise. Higher values of PSNR indicate a better reconstruction. SSIM index quantifies the similarity between two images by leveraging the dependencies that exist among adjacent pixels. It assesses the structural integrity within images through a methodical evaluation across various locations using a moving window approach. Higher values of SSIM indicate a better reconstruction. We compute NMSE, PSNR, and SSIM using the following formulas:

$$\text{NMSE}(\hat{\mathbf{v}}, \mathbf{v}) = \frac{\|\hat{\mathbf{v}} - \mathbf{v}\|_2^2}{\|\mathbf{v}\|_2^2} \quad (5)$$

$$\text{PSNR}(\hat{\mathbf{v}}, \mathbf{v}) = 10 \log_{10} \frac{\max(\mathbf{v})^2}{\text{MSE}(\hat{\mathbf{v}}, \mathbf{v})} \quad (6)$$

$$\text{SSIM}(\hat{\mathbf{m}}, \mathbf{m}) = \frac{(2\mu_{\hat{m}}\mu_m + c_1)(2\sigma_{\hat{m}m} + c_2)}{(\mu_{\hat{m}}^2 + \mu_m^2 + c_1)(\sigma_{\hat{m}}^2 + \sigma_m^2 + C_2)} \quad (7)$$

Here, $\hat{\mathbf{v}}$ is the reconstructed MRI volume, $\mathbf{v}$ is the ground truth MRI volume, and $\hat{\mathbf{m}}$ and $\mathbf{m}$ are patches in the reconstructed and ground truth images, respectively. All definitions and parameters follow the standard practice (Zbontar et al., 2019).

## 3.8 Reconstruction Models for Comparative Evaluations

To assess the influence of our proposed contrastive learning pretraining framework on MRI reconstruction, we tested the reconstruction performance using four state-of-the-art DL MRI reconstruction models from the literature, namely U-Net (Ronneberger et al., 2015), D5C5 (Schlemper et al., 2018), MICCAN (Huang et al., 2019) and ReconFormer (Guo et al., 2024). We trained these reconstruction models using the undersampled images without contrastive learning pretraining (w/o CL-MRI) and using the latent representations from contrastive learning pretraining (w/ CL-MRI), and compared the performances quantitatively and qualitatively.

# 4 Experiments

In this section, we present the main experiments that we conducted in order to assess the proposed framework.

**Experiment 1: Reconstruction Accuracy**

In this experiment, we compared the reconstruction performance on 49 MRI volumes (392 2D axial slices) from the FLAIR dataset described in Section 3.5. We evaluated slice-by-slice reconstruction performance in terms of NMSE, PSNR, and SSIM under acceleration factors of 4X, 6X, 8X, 10X, and 12X and presented in Fig. 2 (average and standard deviation across all slices are presented). We also illustrate the reconstructions of a single 2D MRI slice under an acceleration factor of 8X in order to show the qualitative improvements extended by our proposed contrastive learning framework.

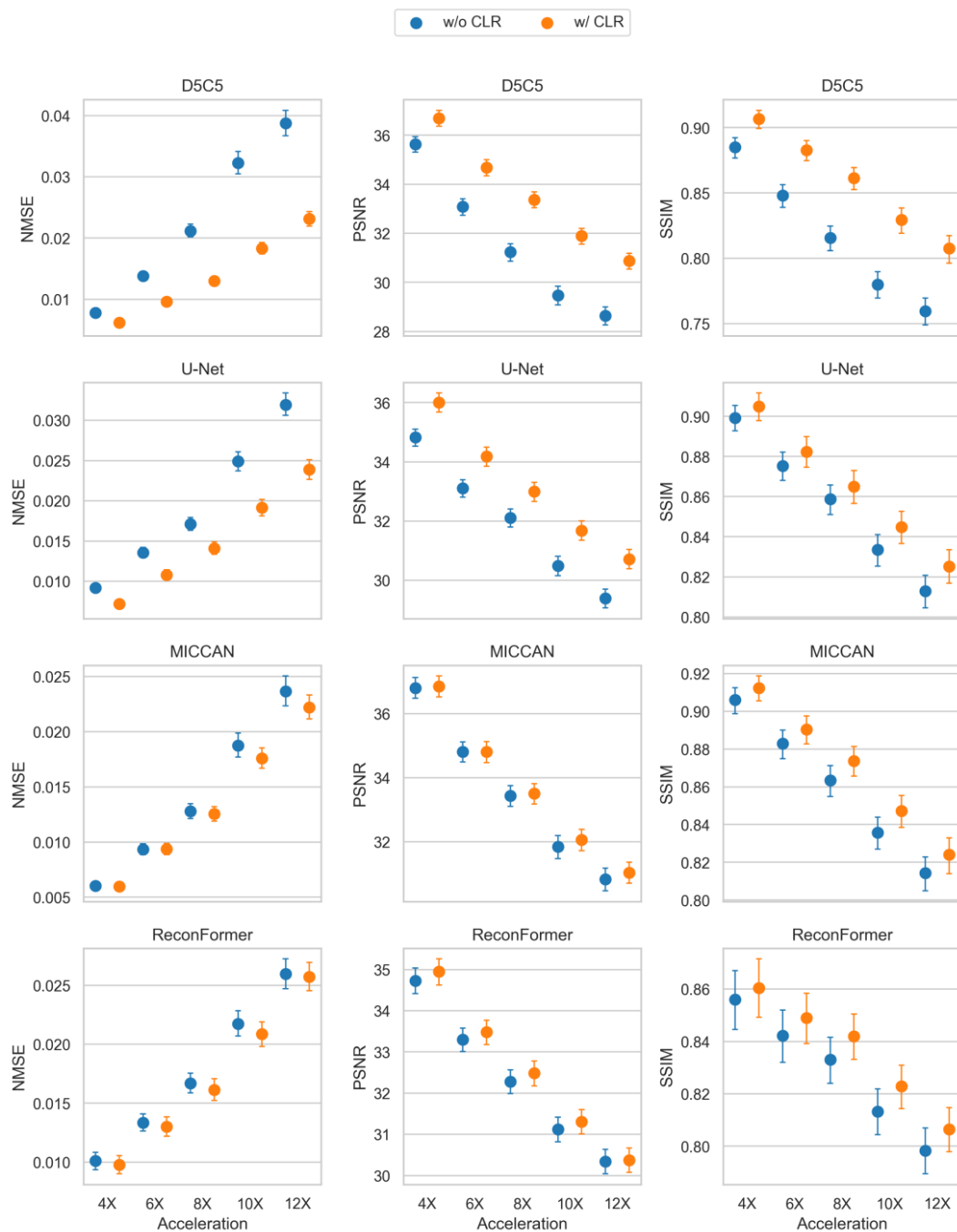

Fig. 2: Quantitative evaluation of reconstruction performance under different acceleration factors for different reconstruction models with (w/ CL-MRI) and without (w/o CL-MRI) contrastive learning.

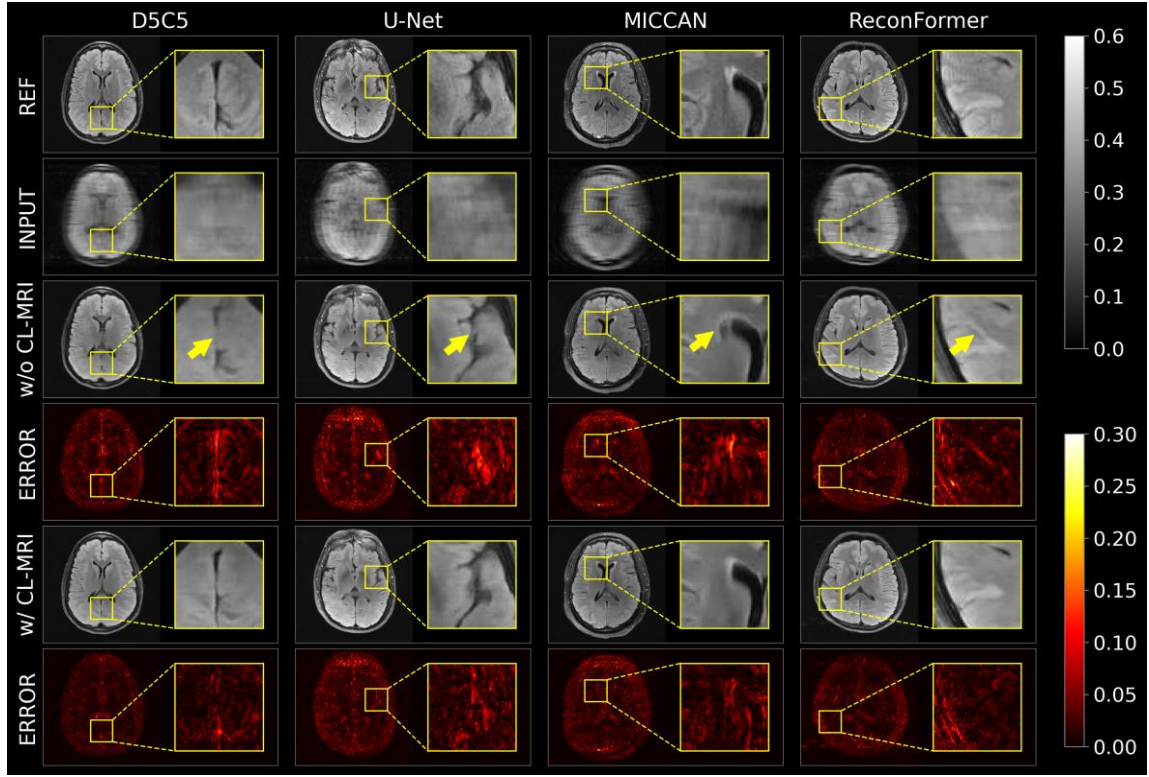

Fig. 3: Qualitative evaluation of reconstruction performance under an acceleration factor of 8X for different reconstruction models with (w/ CL-MRI) and without (w/o CL-MRI) contrastive learning.

**Experiment 2: Reconstruction Accuracy (Out-of-distribution)**

In this experiment, we assessed the capabilities of our contrastive learning feature extraction framework that was trained on brain MRI data, to boost the reconstruction performance of knee data described in Section 3.5. It is worthwhile noting that the contrastive learning features extraction model was trained purely using brain data and we did not use any knee data to retrain or finetune the model weights. The knee data was only utilized to train the D5C5 reconstruction models and for inference. We evaluated slice-by-slice reconstruction performance in terms of NMSE, PSNR, and SSIM at acceleration factors of 4X, 6X, and 8X and presented in Table I (average and standard deviation across all slices are presented). We also illustrate the D5C5 reconstructions of 2D knee MRI slices under acceleration factors of 4X, 6X, and 8X in order to show the qualitative improvements extended by CL-MRI in Fig. 4.

Table I. Quantitative evaluation of reconstruction performance under different acceleration factors for the D5C5 reconstruction model with (w/ CL-MRI) and without (w/o CL-MRI) contrastive learning for the fastMRI PD knee data.

| Acceleration | Metric | w/o CL-MRI | w/ CL-MRI |
|---|---|---|---|
| 4X | NMSE (↓) | 0.0052 ± 0.0027 | 0.0049 ± 0.0025 |
| | PSNR (↑) | 38.09 ± 2.47 | 38.26 ± 2.37 |
| | SSIM (↑) | 0.9200 ± 0.0347 | 0.9224 ± 0.0326 |
| 6X | NMSE (↓) | 0.0106 ± 0.0060 | 0.0098 ± 0.0050 |
| | PSNR (↑) | 35.01 ± 2.72 | 35.27 ± 2.69 |
| | SSIM (↑) | 0.8817 ± 0.0485 | 0.8862 ± 0.0444 |
| 8X | NMSE (↓) | 0.0178 ± 0.0098 | 0.0164 ± 0.0073 |
| | PSNR (↑) | 32.76 ± 2.50 | 32.95 ± 2.44 |
| | SSIM (↑) | 0.8492 ± 0.0522 | 0.8545 ± 0.0489 |

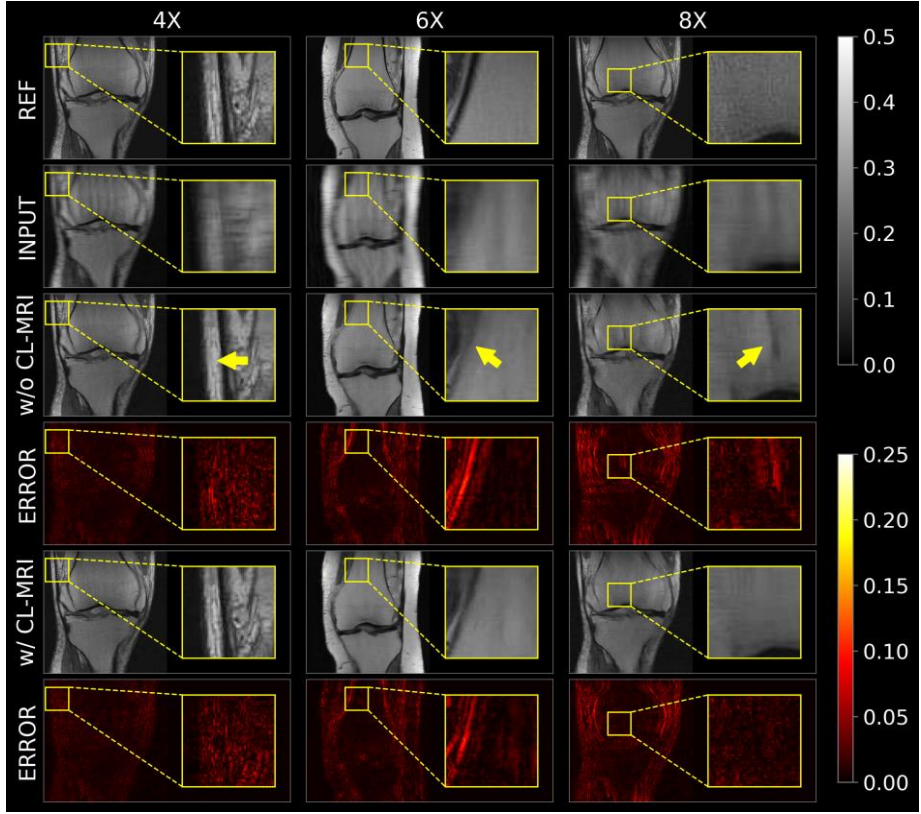

Fig. 4: Qualitative evaluation of reconstruction performance under different acceleration factors for the D5C5 reconstruction model with (w/ CL-MRI) and without (w/o CL-MRI) contrastive learning for the fastMRI PD knee data.

**Experiment 3: Robustness to Sampling Mask**

In this experiment, we assessed the robustness of CL-MRI to a change in the k-space undersampling. Note that all model components, i.e. the contrastive learning feature extraction model and the D5C5 reconstruction models were trained under a random 1D cartesian sampling pattern, whereas in this experiment we changed the mask to an equispaced sampling pattern during inference. We evaluated slice-by-slice reconstruction performance in terms of NMSE, PSNR, and SSIM at an acceleration factor of 4X as presented in Table II (average and standard deviation across all slices are presented).

Table II. Quantitative evaluation of reconstruction performance with different k-space sampling patterns under an acceleration factor of 4X for D5C5 reconstruction model with contrastive learning.

| Metric | random sampling | equispaced sampling |
| --- | --- | --- |
| NMSE ($\downarrow$) | $0.0062 \pm 0.0030$ | $0.0057 \pm 0.0028$ |
| PSNR ($\uparrow$) | $36.69 \pm 2.29$ | $37.07 \pm 2.33$ |
| SSIM ($\uparrow$) | $0.9067 \pm 0.0506$ | $0.9114 \pm 0.0490$ |

**Experiment 4: Robustness to Measurement Noise**

In this experiment, we assessed the robustness of CL-MRI to measurement noise. We introduced complex additive Gaussian noise to the input k-space measurements (baseline), thereby simulating signal-to-noise ratio (SNR) levels of 40, 35, 30, 20, and 25 dB. Note that all model components, i.e. contrastive learning feature extraction model and the reconstruction models were trained without the introduction of noise, whereas in this experiment we introduce noise at various SNR levels during inference. We evaluated slice-by-slice reconstruction performance in terms of NMSE, PSNR, and SSIM at an acceleration factor of 8X as presented in Fig. 5 (average and standard deviation across all slices are presented).

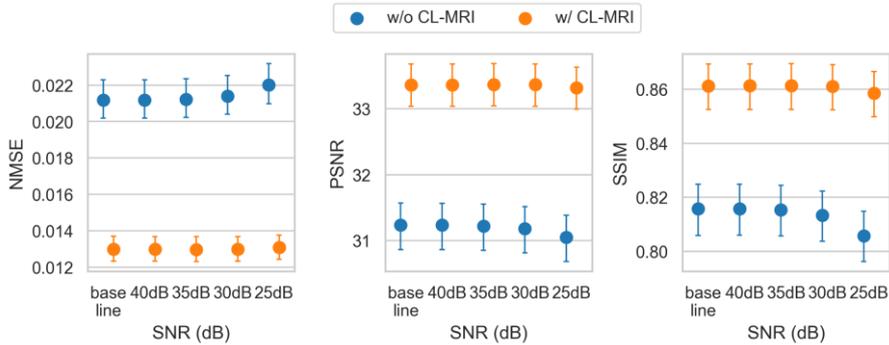

Fig. 5: Quantitative evaluation of reconstruction performance under an acceleration factor of 8X for the D5C5 reconstruction model with (w/ CL-MRI) and without (w/o CL-MRI) contrastive learning in different noise levels.

**Experiment 5: Robustness to Pathology**

Despite significant advancements in accelerated MRI through deep learning, the robustness of under-represented data such as pathology remains a notable concern. Even advanced deep learning models showcased in the fastMRI challenges of 2019 and 2020 struggled to accurately reconstruct pathological abnormalities. To address this challenge, (Zhao et al., 2022) provided subspecialist expert bounding box annotations for different brain pathology categories for the fastMRI brain images. We conducted an experiment to evaluate the reconstruction performance of these annotated slices to assess our framework's resilience to abnormalities. We evaluated slice-by-slice performance for D5C5 reconstructions in terms of NMSE, PSNR, and SSIM at an acceleration factor of 4X and presented in Fig. 6 (average and standard deviation across all slices are presented). We also illustrate three examples of D5C5 reconstructions of three different pathologies (lacunar infarct, resection cavity, and white matter lesion) under acceleration factors of 4X in order to show the qualitative improvements extended by our proposed contrastive learning framework (see Fig. 7).

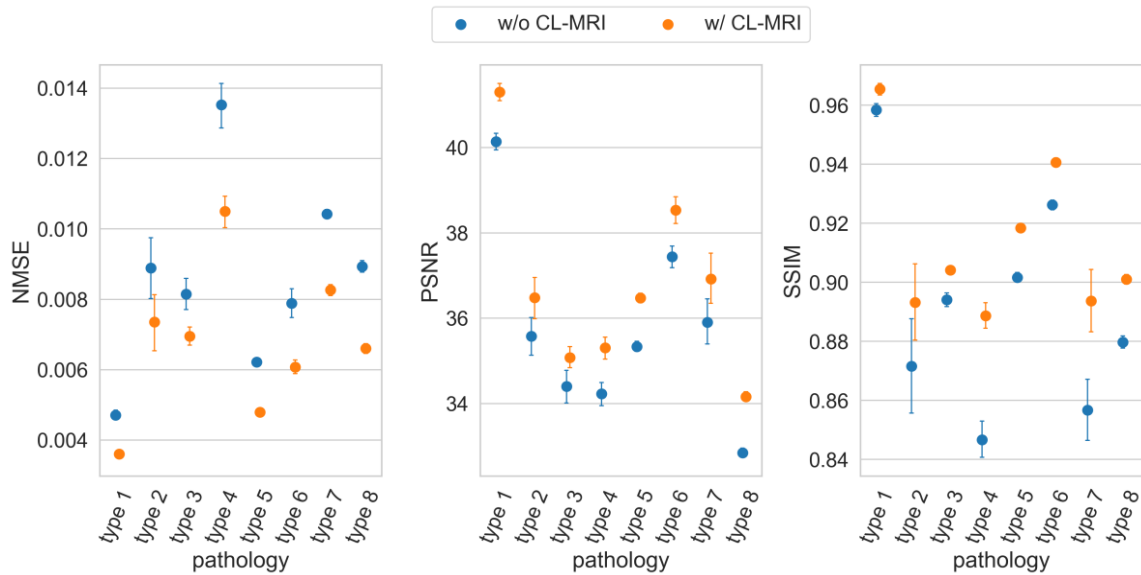

Fig. 6 Quantitative performance comparison on the scans which were identified to contain brain pathology by Zhao et al., (2022) under an acceleration factor of 4X for the D5C5 reconstruction model with (w/ CL-MRI) and without (w/o CL-MRI) contrastive learning. The types of pathologies are type 1: Posttreatment change, type 2: Possible artifact, type 3: Nonspecific white matter lesion, type 4: Enlarged ventricles, type 5: Craniotomy, type 6: Lacunar infarct, type 7: Resection cavity, and type 8: Nonspecific lesion.

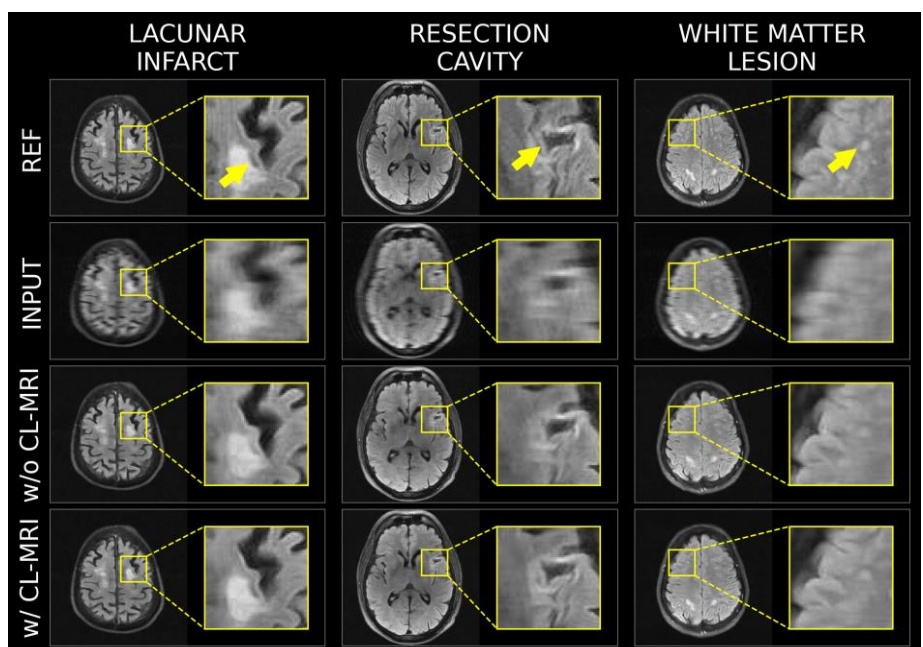

Fig. 7: Qualitative evaluation of reconstruction performance of pathology cases under an acceleration factor of 4X for the D5C5 reconstruction model with (w/ CL-MRI) and without (w/o CL-MRI) contrastive learning.

**Experiment 6 Analysis of latent space – Alignment**

For the Alignment analysis, we used the pretrained contrastive feature extraction network which was trained using brain data under acceleration factors of 2X, 4X, 6X, and 8X, and computed the $\ell_2$-distances between latent features of the positive pairs and observed their proximity. For comparison, we also computed the $\ell_2$-distances between latent features of the positive pairs from a randomly initialized contrastive feature extraction network and observed their proximity. Fig. 8 demonstrates the results in the form of histograms of the distances categorized based on the pair of accelerations utilized.

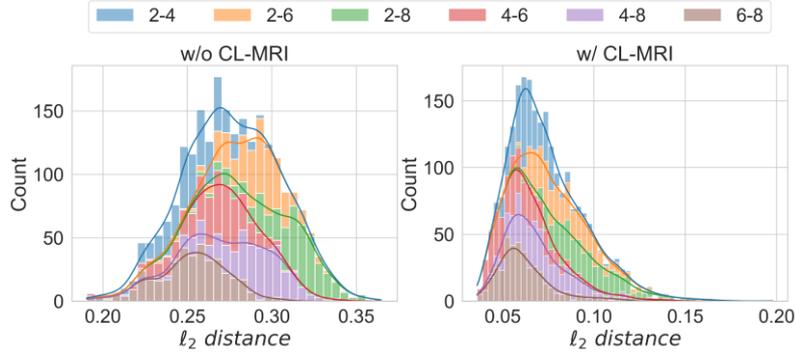

Fig. 8 Distribution of $\ell_2$-distances between the positive pairs obtained from a randomly initialized contrastive feature extraction network (w/o CL-MRI) and a trained contrastive feature extraction network (w/ CL-MRI) under different acceleration factors.

**Experiment 7 Analysis of Latent Space - Uniformity**

For the Uniformity analysis, we used the pretrained contrastive feature extraction network which was trained using brain data under acceleration factors of 2X, 4X, 6X, and 8X. For the convenience of visualization, we plot the contrastive features in a normalized $\mathbb{R}^2$ space. To reduce the dimensionality, we use t-SNE (Maaten and Hinton, 2008) which transforms the similarities among data points into joint probabilities and aims to reduce the Kullback-Leibler divergence between the joint probabilities of the low-dimensional embedding and the original high-dimensional data. For comparison, we also visualize the distribution of extracted contrastive latent features from a randomly initialized contrastive feature extraction network and observed their spread on a normalized $\mathbb{R}^2$ space. Fig. 9 demonstrates the distributions of features with Gaussian kernel density estimation (KDE) on the unit circle.

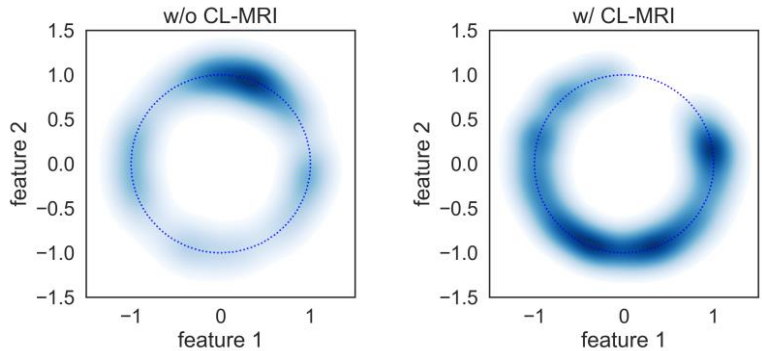

Fig. 9 Gaussian kernel density maps of the reduced features obtained from a randomly initialized contrastive feature extraction network (w/o CL-MRI) and a trained contrastive feature extraction network (w/ CL-MRI).

**Experiment 8 Mutual Information Maximization**

Contrastive learning has a mutual information perspective to it. It can be proven that minimizing the loss function in Eq. (3) will maximize a lower bound on mutual information (Oord et al., 2019). In this experiment, we empirically observed this perspective using acceleration factors of 2X, 4X, 6X, and 8X. We compute the mutual information between the contrastive latent representations and their corresponding fully sampled ground truths and compare them with the mutual information between the undersampled images and their fully sampled ground truths. We plot the results against the reconstruction evaluation metrics, NMSE, PSNR, and SSIM using a D5C5 reconstruction model in order to observe the empirical relationship between the mutual information and reconstruction metrics as shown in Fig. 10.

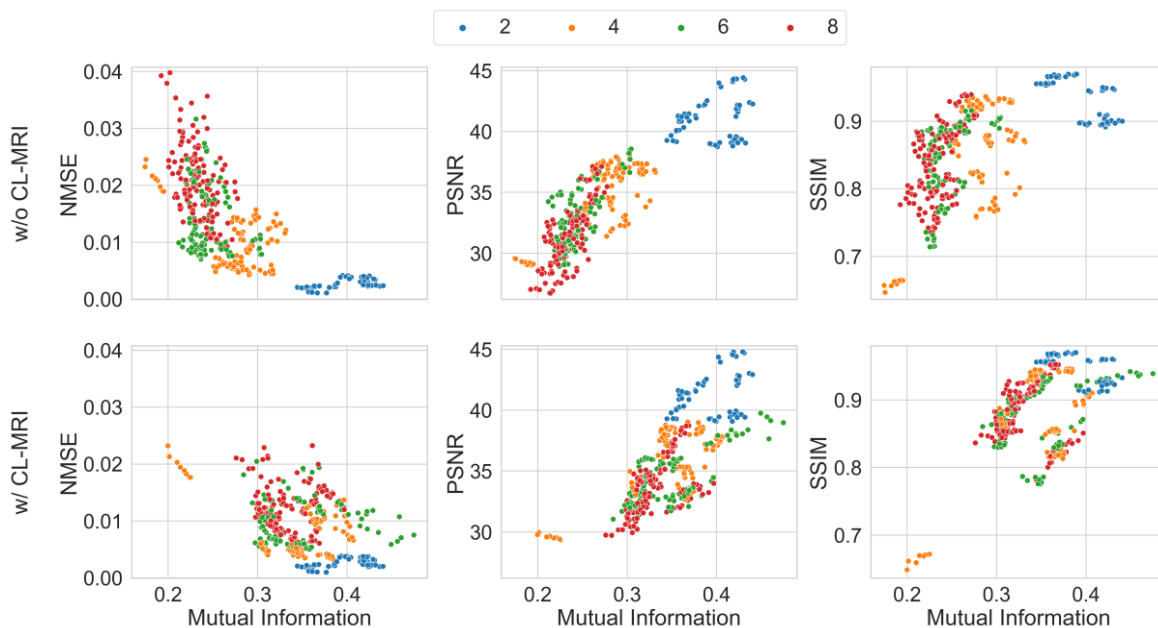

Fig. 10 The variation of different quantitative reconstruction metrics against the mutual information between the input and ground truth for the conventional reconstruction pipeline (w/o CL-MRI) and the proposed contrastive learning framework (w/ CL-MRI) under different acceleration factors for the D5C5 reconstruction model.

**Experiment 9 Model Convergence**

Finally, we analyze the effect of the proposed CL-MRI framework on the DL reconstruction model training convergence. We visualize the training and validation loss curves of D5C5, U-Net, MICCAN, and ReconFormer reconstruction models with and without CL-MRI, and presented in Fig. 11.

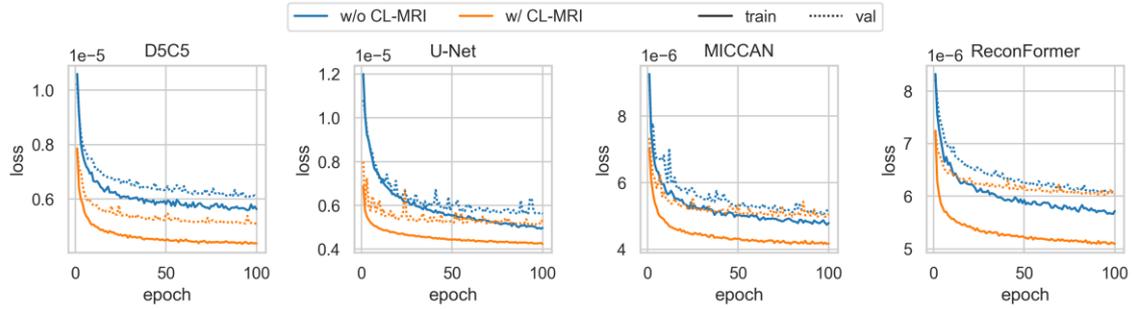

Fig. 11 Training and validation loss curves of D5C5, U-Net, MICCAN, and ReconFormer reconstruction models with (w/ CL-MRI) and without (w/o CL-MRI) contrastive learning.

## 5 Results

The results of our experiments show enhanced reconstruction accuracy across various acceleration factors and datasets, alongside robustness under adversarial conditions and transfer learning capabilities on diverse MRI datasets. Fig. 3 displays the reconstructed images using various methods with and without CL-MRI, at an acceleration factor of 8X on the FLAIR brain images. Visual examination reveals that CL-MRI reconstructions achieve high quality with reduced error and reveal intricate details. The yellow arrow in the D5C5 example clearly points to a case where the D5C5 model fails to reconstruct a continuous feature in the brain image, which is rectified by CL-MRI. Moreover, the yellow arrow in the MICCAN example indicates a case where the reconstructed image contained an artifact, which was eliminated by CL-MRI, thus highlighting its capability to produce more accurate reconstructions. This is further evident in Fig. 2, where reconstructions with CL-MRI show superior NMSE, PSNR, and SSIM values compared to their counterpart reconstructions without CL-MRI across all considered acceleration factors and all reconstruction models. It is also noteworthy that the performance gain increases with the acceleration factor; for instance, the difference in performance of NMSE scores for D5C5 is much larger at 12X acceleration compared to 4X acceleration, indicating the suitability of our proposed CL-MRI framework to enhance performance at very high acceleration factors.

Fig. 4 displays the reconstructed images where the contrastive learning feature extractor was trained on brain data but was used to extract contrastive features from knee data. Visual examination reveals that images reconstructed with CL-MRI achieve high quality with reduced error and intricate details. The yellow arrow of the 8X example indicates a case where a D5C5 reconstruction contained an artifact, that was eliminated by CL-MRI, further highlighting the framework's capability to produce more accurate reconstructions even when the contrastive learning feature extractor was trained on a completely different anatomy. This is further confirmed in Table I, where reconstructions with CL-MRI show superior NMSE, PSNR, and SSIM values compared to their counterpart reconstructions without CL-MRI across all considered acceleration factors.

From Table II, it is clear that CL-MRI is robust to the sampling mask used for k-space undersampling, where only a slight change of performance was observed when an equispaced mask was enforced during inference. Fig. 5 shows that the proposed framework is robust to the measurement noise at SNR levels of 40, 35, 30, 20, and 25 dB where the reconstructions with CL-MRI show superior quantitative metrics at all considered noise levels compared to the reconstructions without CL-MRI.

Fig. 6 clearly shows the robustness of the contrastive learning framework to pathology, where the quantitative reconstruction scores were assessed on the MRI images containing various types of classified pathology, i.e., post-treatment changes, possible artifacts, nonspecific white matter lesions, enlarged ventricles, craniotomy, lacunar infarcts, resection cavities, and nonspecific lesions. In all these types of pathology, the reconstructions with CL-MRI exhibit superior quantitative metrics compared to those without CL-MRI in terms of NMSE, PSNR, and SSIM. Furthermore, Fig. 7 demonstrates three cases where this result was visually confirmed. The first column of Fig. 7 points to a lacunar infarct where the reconstruction with CL-MRI is much sharper and exhibits superior contrasts compared to the reconstruction without CL-MRI. The middle column of Fig. 7 focuses on a resection cavity where the reconstruction without CL-MRI is much more blurred and omits certain intricate details of the ROI compared to the reconstruction with CL-MRI. Finally, the last column of Fig. 7 points to a white matter lesion where the reconstruction with CL-MRI has higher image contrast that permits accurate visualization of the lesion.

In terms of the analysis of the latent space, Fig. 8 demonstrates that the $\ell_2$-distances between all positive pairs have been drastically reduced after contrastive learning pretraining, suggesting that it enforces the alignment of positive pairs with each other. It is also observed from the right panel of Fig. 8 that pairs like 2-4, 4-6, and 6-8 have much smaller $\ell_2$-distances compared to pairs like 2-6, 2-8, and 4-8, indicating that consecutive acceleration factors are much easier to align since they share a greater quantity of mutual information. The KDE maps in Fig. 9 show that without CL-MRI, the latent representations are clustered non-uniformly into certain regions in the latent space, whereas contrastive learning enforces a uniform distribution of latent representations, which is a supporting factor for a successful downstream task according to Wang and Isola, (2020).

Fig. 10 validates the mutual information perspective of the general supervised DL setting. By comparing the top row of panels with the bottom row of panels of Fig. 10, it can be clearly seen that the scatter points have shifted toward the down-right direction for NMSE and shifted toward the up-right direction for PSNR and SSIM. The horizontal movement indicates that the mutual information has increased on the contrastive latent representations compared to the undersampled images with respect to the fully sampled ground truth images, and the vertical movements indicate performance improvement. The combination of these movements empirically suggests that an increment in mutual information at the input aligns with higher reconstruction performance, which validates the general understanding of supervised DL training.

Fig. 11 demonstrates the improved model training and convergence due to CL-MRI. In all four considered models, D5C5, U-Net, MICCAN, and ReconFormer, it can be seen that the training and validation losses are comparatively lower when the model was pretrained using CL-MRI. Additionally, it was observed that the training converged faster with CL-MRI compared to without CL-MRI. Therefore, from a model optimization perspective, our framework allows faster training of reconstruction models, resulting in efficient utilization of computational resources.

# 6   Discussion

Our proposed contrastive learning framework has great potential for improving the reconstruction accuracy of undersampled MRI. The results show not only quantitative performance improvements across various reconstruction models and a range of acceleration factors with respect to NMSE, PSNR, and SSIM but also qualitative enhancements in terms of artifact elimination and the reconstruction of more accurate images with sharper contrasts and reduced blurring. Additionally, the results suggest that the proposed framework is robust to different k-space undersampling masks, measurement noise, as well as underrepresented data in the training dataset such as pathology. We showed that the CL-MRI produces latent spaces that promote properties such as Alignment and Uniformity, which have proven to aid downstream task accuracy. We illustrate the mutual information maximization perspective of contrastive learning and demonstrate its relationship with quantitative reconstruction performance improvements.

The Out-of-Distribution results are significant as they show the transferability of CL-MRI among different datasets with different anatomical regions. The contrastive learning feature extractor trained on brain data was successfully implemented in the reconstruction of knee data, which is particularly important in medical imaging tasks such as undersampled MRI reconstruction due to the scarcity of medical imaging datasets. Furthermore, the proposed contrastive learning framework can be implemented even when fully sampled scans are not available, thereby eliminating the requirement for obtaining fully sampled data, which may be time-consuming and costly.

It is important to highlight that the comprehensive experimental results demonstrate success over a range of downstream DL reconstruction models, which suggests the framework's superior generalizability across DL reconstruction models. Given the rapidly growing literature on DL model architectures for MRI reconstruction, the CL-MRI approach is a viable solution to further improve the reconstruction accuracy of other reconstruction models across a wide range of acceleration factors.

Despite its strengths, we identify several limitations in our proposed CL-MRI framework. One of the major issues is the requirement of large memory for contrastive learning pretraining. As indicated by Algorithm 1, contrastive learning operates on batches of positive pairs. Thus, the batch size for the loss function in Eq. (3) is proportional to the number of acceleration factors used (e.g., if we use four accelerations, i.e., 2X, 4X, 6X, and 8X, the batch size for the loss function becomes four times larger), which will require increased GPU memory to process. Another limitation is that during inference, the contrastive learning phase will require additional computation to generate contrastive features, which would require additional time, memory, and FLOPs. However, given the considerable gains in reconstruction accuracy, it can be considered a reasonable tradeoff.

Regarding contrastive learning algorithms, CL-MRI is fundamentally built upon one of the very primitive algorithms known as SimCLR (Chen et al., 2020). Nonetheless, advancements like Swav (Caron et al., 2020), Moco (He et al., 2020), DINO (Caron et al., 2021), BYOL (Grill et al., 2020), and Barlow Twins (Zbontar et al., 2021) have demonstrated comparable or even superior enhancements in diverse downstream tasks compared to SimCLR. In the future, we plan to explore these techniques in the undersampled MRI reconstruction setting.

# 7   Conclusion

The proposed Contrastive Learning MRI framework utilizes self-supervised contrastive learning to improve the representation of MRI images in the latent space by maximizing mutual information among different undersampled representations and optimizing the information content at the input of the downstream reconstruction task. Our experiments confirm that CL-MRI has the potential to improve the reconstruction accuracy of a wide range of DL reconstruction models and acceleration factors, even when faced with adversarial conditions like noisy measurements, different sampling masks, variable anatomical regions, and pathological abnormalities. Additionally, our analysis of representation quality with a focus on crucial latent space properties including alignment and uniformity, indicated that MRI features are well dispersed within the latent space. In conclusion, CL-MRI constitutes a promising solution to enhance the accuracy of conventional supervised DL-based MR reconstruction.

## Funding

This work was partly funded by the Australian Research Council Discovery Program (DP210101863), and Australian Research Council Fellowship Program (IM230100002).